\documentclass[
  ,draft            
  ]
  {aipproc}

\layoutstyle{8x11double}

\begin{document}

\title{On the evolutionary connection between AGB stars and PNe}


\author{F.~M. Jim\'enez-Esteban}{
  address={Hamburger Sternwarte, Gojenbergsweg 112, D-21029 Hamburg, Germany.}
}

\author{P. Garc\'\i a-Lario}{
  address={Research and Scientific Support Department of ESA, European Space Astronomy Centre, Villafranca del Castillo, Apartado de Correos 50727,  28080 Madrid, Spain.}
}

\author{D. Engels}{
  address={Hamburger Sternwarte, Gojenbergsweg 112, D-21029 Hamburg, Germany.}
}

\begin{abstract}
 
The `O-rich AGB sequence' is a sequence of colours describing the
location of O-rich AGB stars in the IRAS two-colour diagram
[12]--[25]\,vs\,[25]--[60] (See Figure 1). We propose an evolutionary
scenario for this sequence in which all stars, independent of their
progenitor mass, start the AGB phase in the blue part of the `O-rich
AGB sequence' and then evolve toward redder colors, although only the
more massive stars would reach the very end of the `O-rich AGB
sequence'. The sources located in the blue part of the sequence are
mainly Mira variables, whose mean period is increasing with the IRAS
colours. Most of them will evolve into O-rich Type II (and III)
Planetary Nebulae. Part of the stars located in the red part of the
sequence will change their chemical composition from O-rich to C-rich
during their evolution in the AGB phase, and might evolve into C-rich
Type II Planetary Nebulae.  Hot bottom burning may prevent the
conversion to carbon stars of the rest of sources located in the red
part of the sequence and they will end up as N-rich Type I Planetary
Nebulae.
\end{abstract}

\maketitle


\section{Introduction}

The interpretation of the `O-rich AGB sequence' is a key question in
the understanding of AGB stellar evolution. It has been shown by
several authors (e.g. Bedijn 1987) that this sequence reflects the
increase of optical thickness of the circumstellar envelope (CSE) of
AGB stars, the objects with the thinnest CSE being placed in the
bluest part and those with the thickest CSE in the reddest part of the
sequence. However, this can be interpreted in two different ways: i)
as an {\bf evolutionary sequence} (e.g. van der Veen \& Habing 1988):
every O-rich AGB star would start the AGB phase at the blue end of
this sequence, and would later move while increasing its mass-loss
rate until reaching the red extreme at the end of the AGB; ii) as {\bf
mass sequence} (e.g. L\'epine et al. 1995): their different location
would be just a consequence of their different initial mass, which
would determine the mass-loss rate. A third interpretation was
proposed, which is a combination of the previous ones: all O-rich AGB
stars would move towards redder colours as they increase their
mass-loss rate, but only the more massive stars would be able to reach
the reddest end of the sequence (Garc\'\i a-Lario 1992). In addition,
there is not a common opinion on the evolutionary connection between
different classes of AGB stars and different classes of Planetary
Nebulae (PNe).

\section{The sample}


\begin{figure}
  \includegraphics[height=.40\textheight]{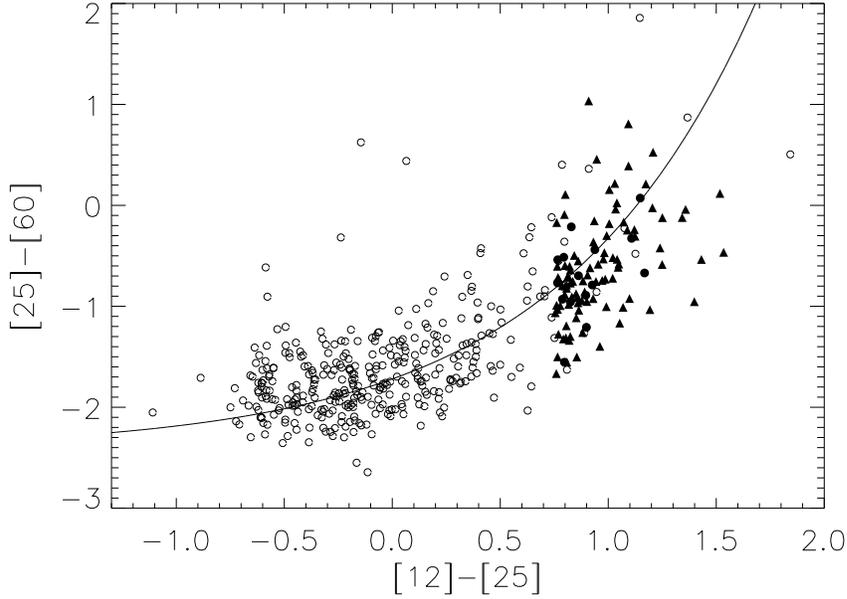}
  \caption{The position of the AGB stars in the sample in the IRAS
 two-colour diagram (open circles for the `Arecibo sample'; filled
 triangles for the `GLMP sample'; filled circles for the few objects
 in common). The solid line is the `O-rich AGB sequence'.}
\end{figure}

Recently, we have published the result of a large scale near-infrared
photometric survey of two different samples of O-rich AGB stars
(Jim\'enez-Esteban et al. 2005a,b): the `Arecibo sample' (363) and the
`GLMP sample' (94). In total we compiled 457 stars providing a good
coverage of the `O-rich AGB sequence' (Figure 1).

The two samples are complementary. The `Arecibo sample' is mainly
formed by objects from the blue part of the sequence, with a small
contribution from very red sources, while the `GLMP sample' contains
exclusively sources in the redder part of the sequence.

\section{RESULTS}

For all the objects in the sample we have collected: J (1.25\,$\mu$m),
H (1.65\,$\mu$m) and K (2.2\,$\mu$m) photometric data from our own
observations or from the 2MASS Point Source Catalogue (PSC); A
(8.28\,$\mu$m), C (12.13\,$\mu$m), D (14.65\,$\mu$m) and E
(21.3\,$\mu$m) flux densities from the MSX6C PSC; and 12, 25, 60 and
100\,$\mu$m\, photometry from the IRAS PSC. Based on these data we
constructed their {\em Spectral Energy Distributions} from the near-
to the far-infrared domain, and estimated the bolometric flux by
integrating, and extrapolating toward both shorter and longer
wavelengths.

The absolute luminosity of AGB stars is still subject to debate. We
have selected 41 sources detected in the direction of the Galactic
Bulge (GB) and assumed a common distance to all these sources
equivalent to the generally assumed distance to the Galactic Center of
$\approx$\,8\,kpc (Reid 1993). Figure 2 shows the distribution of
absolute luminosities. Although a wide range of luminosities is found,
however, the distribution is strongly peaked around
3\,500\,L$_{\odot}$. This maximum is in very good agreement with the
one previously determined by other authors, who studied mainly bluer
samples of AGB stars in the GB (e.g. Habing et al. 1985; Blommaert et
al. 1998; Jackson et al. 2002) and in the solar neighborhood (Knauer
et al. 2001). It appears that the luminosity distribution is similar
throughout the Galaxy and not very dependent on the
colours. Therefore, we assumed a common and constant luminosity of
3\,500\,L$_{\odot}$ for all the sources as a first approximation to
their real luminosity, to obtain the distance and then the galactic
height $z$.

\section{Interpretation of the `O-rich AGB sequence'}

The position of each star in the IRAS two-colour diagram has been
re-defined, via the following parameterization of the `O-rich AGB
sequence':

\begin{center}
$ \left. \begin{array}{l}
$[12]$-$[25]\,=\,0.912\,Ln\,$\lambda \\
\\
$[25]$-$[60]\,=\,$-$2.42\,+\,0.72\,$\lambda
\end{array} \right \} $
\end{center}

\begin{figure}
  \includegraphics[height=.25\textheight]{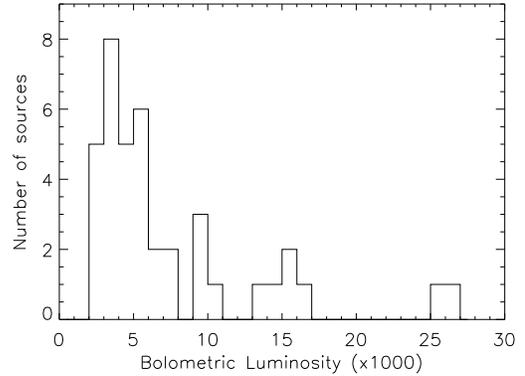}
  \caption{Absolute bolometric luminosity in thousands of L$_{\odot}$
	of the AGB stars belonging to the GB.}
\end{figure}

This way each star was assigned a $\lambda$
value which corresponds to the nearest point on the `O-rich AGB
sequence'. Thus, low values of $\lambda$ correspond to objects
located in the blue part of the sequence and high values of $\lambda$
to those located in the red part of it.

\begin{figure}
  \includegraphics[height=.25\textheight]{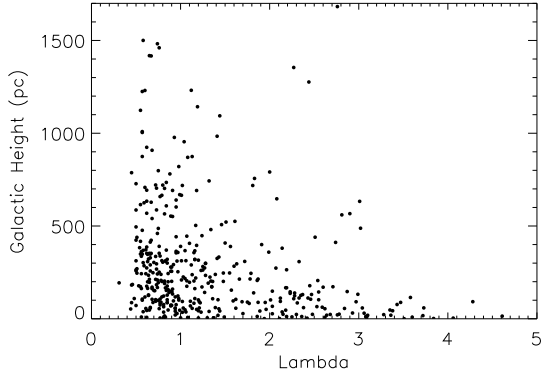}
  \caption{$|z|$ distribution as a function of $\lambda$.}
\end{figure}

Figure 3 shows the distance to the galactic plane ($|z|$) as a
function of the $\lambda$, excluding the GB subsample.  We found a
very clear trend in the sense that sources with low values of
$\lambda$ have a wider distribution than the redder sources.  This
implies that the redder part of the `O-rich AGB sequence' must be
populated mainly with objects of higher mass.

\begin{figure}
  \includegraphics[height=.25\textheight]{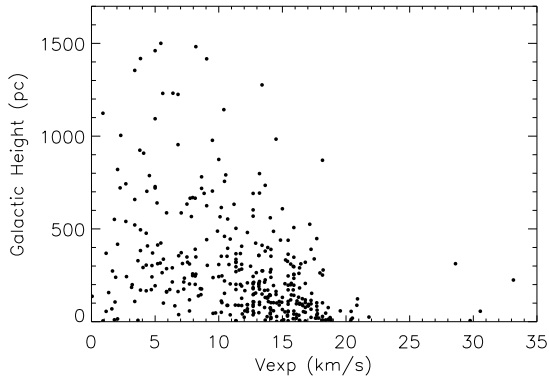}
  \caption{$|z|$ distribution as a function of v$_{exp}$.}
\end{figure}

The expansion velocity (v$_{exp}$) of the circumstellar shell has been
proposed to be correlated with the progenitor mass (Baud \& Habing
1983; Garc\'{\i}a-Lario 1992). v$_{exp}$ was derived for the stars in
our sample from the OH maser measurements. Figure 4 shows that there
is also a very clear correlation between $z$ and v$_{exp}$ in the
sense that there is a deficit of stars with high v$_{exp}$ at high
$z$. In addition, among the sources with low $z$ most of them have
high v$_{exp}$. In Figure 5 we show that sources with high v$_{exp}$
are found in all areas of the `O-rich AGB sequence'. This means that
the objects with high v$_{exp}$, regardless the colour, in mean must
have higher progenitor masses than those with low v$_{exp}$.

\begin{figure}
  \includegraphics[height=.25\textheight]{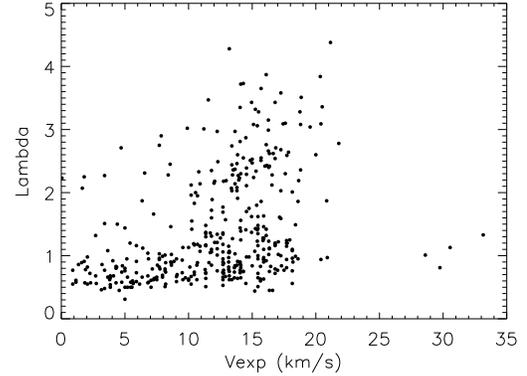}
  \caption{$\lambda$ distribution as a function of v$_{exp}$.}
\end{figure}

We conclude that the red part of the `O-rich AGB sequence' (high
$\lambda$) is mainly populated by sources with higher progenitor mass.
In contrast, the blue part of the `O-rich AGB sequence' is populated
by a combination of AGB stars of low and high progenitor mass.  As
objects with low progenitor mass are almost not found at high values
of $\lambda$, they must abandon the AGB phase without reaching the red
end of the `O-rich AGB sequence'.

\section{Evolutionary connections}

\begin{figure}
  \includegraphics[height=.25\textheight]{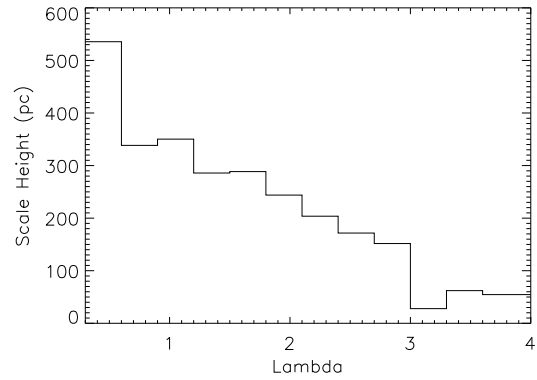}
  \caption{Scale height distribution $H$ as a function of  $\lambda$.}
\end{figure}

In order to search for evolutionary connections between the objects
populating the `O-rich AGB sequence' and other samples of AGB stars
and PNe taken from the literature, we have divided the sample in bins
of $\lambda$ and obtained their galactic scale heights $H$ (Figure 6).
The scale heights decrease with increasing $\lambda$, with particular
large $H$ at the blue end and very low $H$ at the red end. We have
split our sample first into 3 subsamples according to the optical
thickness of their shells (blue, transition, and red) and split
afterwards the blue and red subsamples to discuss the the evolutionary
connection having particular high or low $H$ separately (for details
see Jim\'enez-Esteban et al. 2005c). The resulting subsamples are
listed in Table\,1.


\begin{table}
\caption{Subsamples identified.}

\begin{tabular}{lcrrc}
\hline\noalign{\smallskip}
Sample & $\lambda$ & N of   &  $\approx$H   & Optical     \\
       & range     & stars  & [pc]  & thickness \\
\noalign{\smallskip}\hline\noalign{\smallskip}
Ext. Blue & \multicolumn{1}{r}{$\lambda$\,$\le$\,0.6} &  43  & 550  & Thin          \\
Blue      & 0.6\,$<$\,$\lambda$\,$\le$\,1.2           & 193  & 350  & Thin          \\
Trans.    & 1.2\,$<$\,$\lambda$\,$\le$\,1.8           &  58  & 300  & Thin \& Thick \\
Red       & 1.8\,$<$\,$\lambda$\,$\le$\,3.0           &  66  & 200  & Thick         \\
Ext. Red  & \multicolumn{1}{l}{3.0\,$<$\,$\lambda$}   &  22  &  50  & Thick         \\
\noalign{\smallskip}\hline 
\end{tabular} 
\end{table}

The {\em extremely blue} subsample has an associated galactic scale
height $H$ remarkably similar to that found for optically bright Mira
variables with short periods ($<$\,300$^{d}$; Jura 1994). O-rich Type
III PNe (very low mass O-rich PNe) also have a similar mean $z$
(Maciel \& Dutra 1992).

The {\em blue subsample} shows a value of $H$ similar to that
found by Ortiz \& Maciel (1996) for their sample of OH/IR stars in the
solar neighborhood, and with that obtained by Wood \& Cahn (1977) for
their sample of Mira variables near the Galactic Plane, which is a combination
of both short and long period Miras. Type II PNe, which are a mixture of
O-rich and C-rich PN, have also similar mean $z$ (Maciel \& Dutra 1992).
                                                                                
The {\em transition subsample} populates the region in which AGB stars
become optically thick.  This subsample has an associated $H$ which is
consistent with the value obtained by Jura \& Kleinmann (1992) and
Jura et al. (1993) for intermediate- and long-period
(300\,--\,500$^{d}$) Mira variables, and by Groenewegen et al. (1992)
for a small group of low mass optically bright carbon stars.

We conclude that these first three subsamples contain Mira variables,
with periods increasing with the value of $\lambda$. They mostly will
evolve into O-rich Type II (and III) PNe. A fraction of the stars in
the transition subsample might evolve into C-rich Type II PNe.

The value of $H$ in the {\em red subsample} is similar to the one
obtained for carbon star samples which include a significant fraction
of very red carbon stars (`infrared carbon stars'). (Claussen et
al. 1987; Groenewegen et al. 1992). It is also similar to the mean
value of $z$ associated to N-rich Type I PN (Maciel \& Dutra 1992). We
therefore suspect that part of the stars in the red subsample change
their chemical composition from O-rich to C-rich during their
evolution in the AGB phase as a consequence of the dredge-up processes
that the theory predicts for intermediate mass stars. Hot bottom
burning (HBB) may prevent the conversion to carbon stars of the rest,
ending up as N-rich Type I PN.

The last subsample, the {\em extremely red subsample}, has a extremely
low value of $H$. These objects should then represent the most massive
AGB stars in our Galaxy. These stars must be massive enough to
activate HBB. Note that there is no PN class which can be associated
to this subsample. It is possible that these massive stars evolve so
fast that when the central star reaches the necessary temperature to
ionize the envelope, the shell is still thick enough to prevent its
detection as a PN in the optical. These massive AGB stars may never
form visible PNe. Rather they would evolve as `infrared PNe' and might
be related with the rare group of so-called `OHPNe' (Zijlstra et
al. 1989), heavily obscured sources which show both OH maser and radio
continuum emission.


\begin{theacknowledgments}

This work was partially supported by the Spanish Ministerio de Ciencia
y Tecnolog\'\i a through grant AYA2003$-$09499. This research has made
use of the SIMBAD database, operated at CDS, Strasbourg, France. This
publication makes also use of data products from the Two Micron All
Sky Survey.

\end{theacknowledgments}


\end{document}